\documentclass{article}

\usepackage{amsmath}
\usepackage{amsfonts}
\usepackage{amsmath}
\usepackage{latexsym}
\usepackage{graphicx}
\DeclareMathOperator{\rot}{rot}

\begin{document}

\title{Radiation of Relativistic Particles in a Quasi-Homogeneous
Magnetic Field}

\author{V.~Epp\footnote{e-mail: epp@tspu.edu.ru},
T.G.~Mitrofanova\\
Tomsk State Pedagogical University, 634041 Tomsk, Russia}
\date{}
\maketitle
\begin{abstract}
Spectrum of radiation of a relativistic particle moving in a
nonhomogeneous magnetic field is considered. The spectrum
depends on the pitch-angle $\alpha$ between
the velocity direction and a line tangent to the field line.
In case of very small $\alpha$ the particle generates so-called
curvature radiation, in an intermediate case undulator-kind radiation
is produced. In this paper we present the calculations of radiation
properties in a case when both curvature and undulator radiation is
observed.
\end{abstract}

It is well known that the spectrum of radiation of a relativistic
particle depends on the pitch-angle $\alpha$ between the velocity
direction and a line tangent to the field line. One can distinguish
three specific cases

1. $\alpha \gg\gamma^{-1}$   synchrotron radiation,

2. $\alpha\leq\gamma^{-1}$     undulator radiation,

3. $\alpha\ll\gamma^{-1}$      curvature radiation,

\noindent
where $\gamma=(1-v^2/c^2)^{-1/2}$, $v$ is the particle velocity, $c$ is
the speed of light.

Radiation of the particle in the first case  is assumed to be
synchrotrone one with the radius of the orbit equal to the radius of a
helix. The maximum in the radiation spectrum falls on high number
of harmonics.

The radiation in the second case is formed along all path of the
particle and is assumed to be of undulator type, when the main part of
radiation power is emitted in some few first harmonics. The third case
is usually realized in a strong magnetic field when the orthogonal
component of velocity vanishes quickly due to synchrotron radiation and
the particle moves almost along the field line. This approximation is
usually accepted for the particles in the pulsar magnitosphere.
Corresponding radiation is called curvature radiation. General formula
for the cases 1 and 3 were obtained in refs \cite{zh,ch}. The
principle point of these papers is that the equations of motion were
expanded in power series with respect to time $t$ in the vicinity of
fixed moment $t=0$. But if $\alpha \leq\gamma^{-1}$, the radiation
spectrum is formed along rather great part of the particle path. This
part can include several loops of the helix. In this case one can
expand in a series only slowly varying functions in equation of
motion. The rapidly oscillating part of motion produces radiation of
undulator type.

In this paper we present the calculations of radiation properties in a
case when both curvature and undulator radiation is observed.

In a general case the curved magnetic field line can be approximated
with an arc of a circle. If the magnetic field does not depend on the
coordinate orthogonal to the plane of magnetic line, the field
$\vec H(\vec r)$ can be represented in a form $H_r=0$,
$H_{\varphi}=H(r)$, $H_z=0$, where $r$,$\varphi$,$z$ are the
cylindrical coordinates.

It follows immediately from the condition $\rot\vec H=0$, that
$H(r)=I/r$, where $I=const$. This is the field of a straight current or
the field of toroidal solenoid in its inner part.

Let us consider equations of motion of a charged particle in such
field. It is easy to find three first integrals of motion: energy,
axial momentum and z-component of momentum. Thus, we can rewrite the
equations of motion in terms of first-order equations
\begin{eqnarray}
\dot{r}^2&=&V^2-U(r),\\ \nonumber
\dot{\varphi}&=&V_{0\varphi}\frac{r_0}{r^2},\\
\dot{z}&=&V_{0z}+{\cal E}c\ln\frac{r}{r_0},\nonumber
\end{eqnarray}

\noindent
where $V=const$ is the particle velocity, $V_{0\varphi}$
and $V_{0z}$ are axial and z-components of velocity
respectively, $r_0$ is the initial coordinate, ${\cal
E}=eI/mc^2\gamma$, $e$ and $m$ are the charge and mass of the particle.

The function
\begin{eqnarray*}
U(r)=\left(V_{0z}-{\cal E}c\ln\frac{r}{r_0}\right)^2+\frac{V_{0\varphi}^2r_0^2}{r^2}
\end{eqnarray*}
 plays the role of effective potential energy. It has the only minimum
at $r=r_m$, which is given by equation
\begin{eqnarray}
r_m^2\left(V_{0z}+{\cal E}c\ln\frac{r_m}{r_0}\right)=\frac{
V_{0\varphi}^2r_0^2}{{\cal E}c}.
\end{eqnarray}

Solution of eqs (1) gives a great variety of trajectories, which
are rather complicated. We are interested only in motion in a
quasi-uniform magnetic field. It means that the particle path lies
in a small interval
$\bigtriangleup{r}\ll r_0$.  Thus, we consider the particle motion in
the vicinity of minimum of effective potential energy. If the initial
state is $r_0=r_m$, $\stackrel{\cdot}{r_0}=0$, then the solution of eqs
(1) is
\begin{eqnarray}
r(t)=r_0,\;\;\;\varphi(t)=\frac{V_{0\varphi}}{r_0}t,\;\;\;z(t)=V_{0z}.
\end{eqnarray}

Note that this solution is valid for arbitrary $r_0$. It follows from
eq.(2) that the only condition realizing the motion of eq.(3) is that
\begin{eqnarray}
V_{0z}=\frac{V_{0\varphi}^2}{{\cal E}c},\;\;\;V_{0r}=0. \nonumber
\end{eqnarray}

In this case the particle moves along a helix with radius $r_0$ and
constant angle $\eta$ between the magnetic field line and velocity
direction.
\begin{eqnarray}
\tan\eta=\frac{V_{0\varphi}}{{\cal E}c}.
\end{eqnarray}

It is usually assumed in astrophysical applications that after the
particle radiates out all the transversal energy, it moves along the
magnetic field line. Eq. (4) shows that it is true only if $V_{0\varphi}
\ll{\cal E}c$. Generally speaking, the angle $\eta$ can
significantly differ from zero.

We assume further that the magnetic field is quasi-uniform, i.e. ${\cal
E}\gg1$ and $\eta\ll1$. Expanding the function $U(r)$ in a power series
with respect to small $(r-r_m)/r_m$ we obtain the following solution of
eqs (1)
\begin{eqnarray}
r&=&r_m+a\sin(\omega_0t+\delta),\\ \nonumber
\varphi&=&\frac{V_{0\varphi}}{r_m}t,\\
z&=&a\cos(\omega_0t+\delta),\nonumber
\end{eqnarray}

\noindent
with $a=r_m\sqrt{V^2-V_{0\varphi}^2}\,/{\cal E}c$,
$\omega_0={\cal E}c/r_m$ and $\delta$ is an arbitrary initial phase.
According to eqs (5), the particle moves along a curved helix.

Let us calculate the spectral and angular distribution of radiation. We
start with a well known formulae \cite{lan}
\begin{eqnarray}
\frac{d{\cal E}_j}{d\Omega d\omega}=\frac{cR^2}{4\pi^2}|E_j(\omega)|^2,
\end{eqnarray}
\begin{eqnarray}
E_j(\omega)=\frac{ei\omega}{cR}e^{ikR}\int\limits_{-\infty}^\infty
\beta_j(t)e^{i(\omega t-\vec k\vec r)}dt,
\end{eqnarray}

\noindent
where $\beta_j(t)=(\vec V\vec e_j)/c$,
$\vec e_j$ are the unite vectors of polarization. Let
the wave vector $\vec k$ lay in the coordinate pane $yz$ and denote by
$\chi$ the angle between $\vec k$ and axis $y$. Then
$\vec e_\sigma=(-1,0,0)$,  $\vec e_\pi=(0,\sin\chi,\cos\chi)$.

We assume that the particle is ultrarelativistic one and the pitch-
angle satisfies the inequality $\alpha\ll\gamma^{-1}$. This allows us
to expand expression (7) in a power series with respect to small
$\frac{V_{0\varphi}}{r_0}t$, but keeping unexpanded functions of
$\omega_0t$. As a result we obtain
\begin{eqnarray}
\omega t-\vec k\vec r&=&\frac{\omega
t}{2}(\gamma^{-2}+\chi^2)+\frac{\omega V_{0\varphi}}{6r_0}t^3,\\
\beta_{\sigma}(t)&=&\frac{V_{0\varphi}}{r_0}t-\frac{a\omega_0}{c}
\cos(\omega_0t-\delta),\\
\beta_\pi(t)&=&\chi-\frac{a\omega_0}{c}\sin(\omega_0t-\delta).
\end{eqnarray}
After integration in equation (7) and substitution in eq. (6) we find
\begin{eqnarray}
\frac{d{\cal E}_\sigma}{d\Omega d\omega}&=&A[(1+
\psi^2)^2K_{2\over 3}^2(q)- k\sin\delta
K_{2\over 3}(q)f(v)+\frac{1}{4}k^2f^2(p)],\\ \nonumber \frac{d{\cal
E}_\pi}{d\Omega d\omega}&=&A[\psi^2(1+\psi^2)K_{1\over 3}^2(q)+
k\sin\delta K_{1\over 3}(q)f(v)+\frac{1}{4}k^2f^2(p)], \end{eqnarray}

\noindent
where
\begin{eqnarray*}
A&=&\frac{4e^2\omega_0^2\nu^2r_0^2}{3c\pi^2V_{0\varphi}^2},
\;\;\;\psi=\gamma\chi,\;\;\;k=\frac{a\omega_0\gamma}{V_{0\varphi}},
\;\;\;\nu=\frac{\omega}{2\gamma^2\omega_0},\\
q&=&\frac{r_0\omega(1+\psi^2)^{3/2}}{3\gamma^3V_{0\varphi}},
\;\;\;N=\frac{\omega_0r_0}{V_{0\varphi}\gamma}.
\end{eqnarray*}

The function $f(p)$ is defined by following expression

\begin{eqnarray*}
f(p)&=&\left\{
\begin{array}{lll}
\frac{\pi}{\sqrt{3}}\sqrt{-\eta}[J_{-1/3}(p)+J_{1/3}(p)], &
p=\frac{2}{3} \nu N(-\eta)^{3/2}, & \eta\leq 0,\\
\sqrt{\eta}K_{1/3}(p),& p=\frac{2}{3}\nu N\eta^{3/2},& \eta\geq 0,
\end{array}\right.
\end{eqnarray*}
\begin{eqnarray}
\eta=1+\psi^2-\nu^{-1}.\nonumber
\end{eqnarray}
We see that equations (11) consist of three terms. The first gives the
typical synchrotron radiation emitted from an arc of a circle  of
radius $r_0$. Thus, it is a curvature radiation. The third term is
proportional to transversal part of the particle velocity
$V_{\perp}=a\omega_0$.  Parameter $k=V_{\perp}\gamma/V_{0\varphi}$ is
the well known undulator parameter \cite{ud,sr}. Hence, the
third term in eqs (11) represents the undulator radiation. It is
evident that the second term describes some kind of interference of
curvature and undulator radiation.

Let us estimate the characteristic frequencies of each part of
radiation. The main part of curvature radiation is emitted at
frequencies defined by $q\sim 1$, i.e. $\omega\sim\omega_{cr}\sim\frac
{V_{0\varphi}}{r_0}\gamma^3$. The undulator part of radiation is
generated at frequencies, at which $p\sim 1$, i.e.
\begin{eqnarray}
\nu\left|1+\psi^2-1/\nu\right|^{3/2}\sim 1.\nonumber
\end{eqnarray}
This gives
\begin{eqnarray}
\nu\sim\frac{1}{1+\psi^2}\pm\frac{1}{N},\nonumber
\end{eqnarray}

\noindent
thus, $\nu\sim 1$, or $\omega\sim\omega_{ur}\sim\omega_0\gamma^2$. This
means that in adopted assumptions $(k\ll1)$ the undulator radiation
contains only first harmonic of basic frequency $\omega_0$ shifted by
Doppler effect. The ratio $\omega_{ur}/\omega_{cr}\sim N\gg1$ shows that
the curvature and undulator radiation are far separated in spectrum. It
means that even when the intensity of one part of radiation is much
less then another one, we can distinguish the curvature and undulator
radiation.

The intermediate part of radiation, which is given by the second term
in eqs (11) strongly depends on the initial phase $\delta$. If we
observe radiation of an incoherent bunch of particles then this term
should be averaged over $\delta$. As a result this term vanishes.

\begin{figure}[htbp]
\center
\includegraphics[width=76mm]{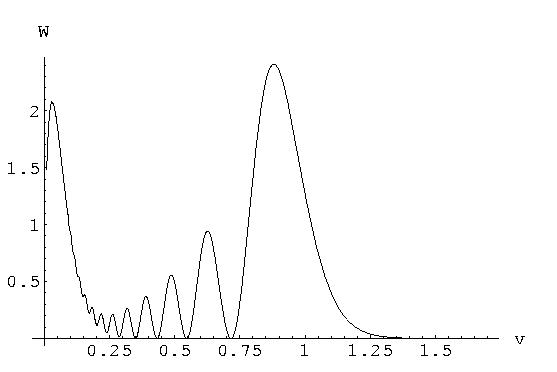}
\caption{Spectrum of radiation for
 $\sigma$-component at an angle $\chi=0$.
The initial phase is $\delta=0$}
\label{fig1}
\end{figure}

\begin{figure}[htbp]
\center
\includegraphics[width=76mm]{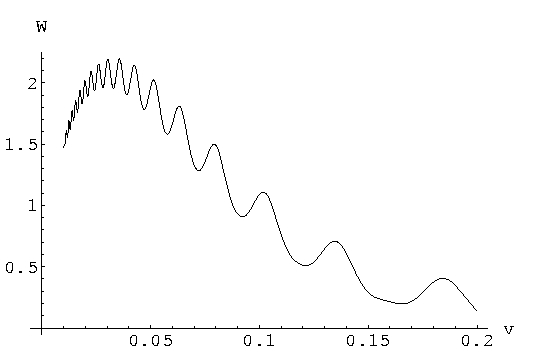}
\caption{Spectrum of radiation for
 $\sigma$-component at an angle $\chi=0$.
The initial phase is $\delta=\pi/6$.}
\label{fig2}
\end{figure}

Figures 1 and 2 show the dependence of spectrum of radiation on
parameters $k$, $\delta$ and $N$. The low-frequency part exhibits the
curvature radiation while the high-frequency part represents the
undulator radiation. We see that the undulator radiation is emitted at
basic harmonic $\nu\approx1$, and the curvature radiation is situated
around frequency $\omega_{cr}\approx\omega_{ur}/N$, i.e.
$\nu\approx0.05$. Figure 2 demonstrates the influence of initial phase
upon the shape of the spectrum. The small oscillations of the spectrum
curve is highly dependent on the value of the initial phase $\delta$.

The dependence on the pitch-angle $\alpha$ is included in the undulator
parameter $k$. Formula (11) shows that the undulator part of radiation
increases with increasing $\alpha$.

\end{document}